\input{epsf}

 \def\CQG{{\it Class. Quantum
Gravity} }
    \def\IJMP{{\it Int. J. Mod. Phys.} }

  \def\PR{{\it
Phys. Rev.} } \def\PRL{{\it Phys. Rev. Lett.} }

\def\al{\alpha}   
\def\ep{\epsilon}   
\def\th{\theta}

 \def\frac#1#2{{\textstyle{{#1}\over
{#2}}}} 
\def\lsim{\mathrel{\rlap{\lower4pt\hbox{\hskip1pt$\sim$}}
\raise1pt\hbox{$<$}}}
\def\gsim{\mathrel{\rlap{\lower4pt\hbox{\hskip1pt$\sim$}}
\raise1pt\hbox{$>$}}} \def\sqr#1#2{{\vcenter{\vbox{\hrule
height.#2pt \hbox{\vrule width.#2pt height#1pt \kern#1pt \vrule
width.#2pt} \hrule height.#2pt}}}}

\def\beq{\begin{equation}} \def\eeq{\end{equation}}
\def\beqa{\begin{eqnarray}} \def\eeqa{\end{eqnarray}}

\def\@fnsymbol#1{\ifcase#1\hbox{}\or *\or \dagger\or \ddagger\or \mathchar "278\or \mathchar "27B\or \|\or **\or \dagger\dagger \or \ddagger\ddagger \or \mathchar"27C \else\@ctrerr\fi\relax}

\long\def\symbolfootnote[#1]#2{\begingroup
\def\thefootnote{\fnsymbol{footnote}}\footnote[#1]{#2}\endgroup}

\newcommand{\mr}[1]{\mathrm{#1}}

\documentclass[smallcondensed,final]{svjour3}

\usepackage{graphics}

\journalname{}

\begin{document}

\title{Estimating radiative momentum transfer through a thermal analysis of the Pioneer Anomaly}

\titlerunning{Estimating radiative momentum of the Pioneer Anomaly}

\author{O. Bertolami \and F. Francisco \and P. J. S. Gil\and J. P\'aramos}

\institute{\at Departamento de F\'{\i}sica, Instituto Superior T\'ecnico;\\ also at Instituto de Plasmas e Fus\~ao Nuclear \and O. Bertolami \at \email{orfeu@cosmos.ist.utl.pt} \and J. P\'aramos (Speaker)\at \email{jorge.paramos@ist.utl.pt}  \and \at Departamento de Engenharia Mec\^anica, Instituto Superior T\'ecnico; \\also at Centro de Ci\^encias e Tecnologias Aeron\'auticas e Espaciais \and F. Francisco \at\email{frederico.francisco@ist.utl.pt} \and P. S. J. Gil \at \email{p.gil@dem.ist.utl.pt} \and \at Address: Av.~Rovisco Pais 1, 1049-001 Lisboa, Portugal}

\date{\today}

\maketitle

\begin{abstract}

A methodology based on point-like sources is discussed, enabling a reliable estimate of the acceleration of the Pioneer 10 and 11 probes caused by thermal effects. A sensitivity analysis of the several parameters of the model allows for a clear indication of the possible thermal origin of the so-called Pioneer anomaly.

\keywords{Pioneer anomaly \and thermal effects}

\PACS{07.87.+v \and 24.10.Pa \and 44.40.+a}

\end{abstract}


\section{Introduction}

\subsection{General Background}

A decade ago, an anomalous, sunbound acceleration on the Pioneer 10 and 11
probes was discovered, using two independent code analyses \cite{JPL,Mark}; this acceleration may be characterized as constant, with a magnitude of $a_\mr{Pio} \simeq (8.5 \pm 1.3) \times 10^{-10} ~\mr{m/s^2}$. Early discussions \cite{JPL} argued that the effect cannot be explained in terms of a misestimation of the systematic effects (of thermal nature, or due to electric or magnetic forces, solar radiation and solar wind pressure, mechanical defects or errors in the Doppler tracking algorithms used, {\it etc}.), despite claims otherwise \cite{Lou}.

Amongst these ``conventional'' candidates, one of the prime contenders is the reaction force due to thermal radiation from the main bus compartment and the radiothermal generators (RTGs), either
directly pointing away from the Sun, or reflected by the main antenna dish. Quite obviously, an acceleration due to thermal dissipation should present a secular evolution parallel to the available power in the RTGs, thus decaying at an approximately exponential rate. This said, one must note that Ref. \cite{Mark} shows that the available data may be fitted to such a signature, {\it i.e.} a linear decay with a time constant larger than 70 years: since the half-life of the plutonium source in the RTGs is $\sim88$ years (that should be effectively less due to thermal coupling degradation), this indicates that thermal effects could account for the Pioneer anomaly. This scenario is being extensively scrutinized by groups within the Pioneer collaboration team \cite{team,Toth}, and has been the motivation of the recent study carried out by our group \cite{Lisbon}. A consistent study of secular and spatial trends is of the utmost importance, in order to ascertain possible thermal or engineering origins of the anomalous acceleration. 

The possibility that the gravitational pull from the Kuiper Belt may originate the
anomalous acceleration was investigated in Refs. \cite{JPL,Vieira,Nietokuiper}; it was found that the reported magnitude of the effect would require that this extended object has a mass two orders of magnitude higher than the commonly considered value of $M_{\mr{Kuiper}} = 0.3 M_{\mr{Earth}}$.

The inability to account for the Pioneer anomaly within the realm of conventional physics (see Ref. \cite{Paramos,Reynaud,Moffat} and references therein) has driven several authors to consider a number of new theoretical ideas to explain it. However, proposals arising from new physics must be taken with due care, since an unambiguous description of the anomaly is still lacking. Indeed, the originally available Doppler measurements were conducted at distances that do not
allow for a clear distinction of the direction of the acceleration: in particular, it is not known if the anomalous acceleration points towards the Sun or the Earth, along the line of sight. This information is of paramount importance: an effect pointing
towards the Sun would indicate a gravitational origin (having discarded the far too weak hypothesis of solar wind pressure), while an anomaly directed at our planet would indicate that its origin lies in a modified Doppler effect (perhaps reflecting unknown physics affecting light propagation and causing an effective blue shift), or the incorrect interpretation of Doppler data, possibly due to systematic errors such as mismodeled Earth orientation parameters, incorrect ephemerides estimates, Deep Space Network or software clock drifts. Also, an anomalous acceleration along the spin axis of the spacecraft would show that it is caused by underestimated onboard systematic effects.

If the acceleration is parallel to the velocity vector, this would indicate that the anomaly is a drag effect -- although this does not appear to be physically motivated, as the reported magnitude of the acceleration requires that the probes fly through a medium with a density of order $10^{-19}~\mr{g/cm^3}$ (see, {\it e.g.}, Ref.\ \cite{Vieira}) -- which should be compared with the density of interplanetary dust (arising from
hot-wind plasma \cite{Kimura}) $ < 10^{-24}~\mr{g/cm^3}$, and interstellar dust (directly measured by the Ulysses spacecraft) $ 10^{-26}~\mr{g/cm^3}$. From a theoretical perspective, a modification of geodetical motion arising from extensions of General Relativity could also account
for a velocity dependent anomalous acceleration \cite{zacuto,Lobo}.

Initially regarded with some suspicion, the interest in the Pioneer anomaly has steadily grown, as disclosed by the number of peer-reviewed publications. The characterization of this anomalous acceleration was a major component of the scientific objectives of two mission proposals put forward to the recent
ESA Cosmic Vision 2015-2025 program \cite{missions}; although not considered by the reviewing board, this remains a burning question that may hint at yet unknown aspects of gravitational physics or, at the very least, allow us to increase our understanding of spaceflight dynamics.

\subsection{Previous Work}

\begin{table*}[ht] \begin{center} \caption{Error budget for the
Pioneer 10 and 11, taken from Ref. \cite{JPL}.}

\label{tableJPL}


\begin{tabular}{rlll}
 \hline\hline
   Item & Description of error budget constituents  & Bias~~~~~             & Uncertainty \\
        &                                           & $10^{-8} ~\rm cm/s^2$ & $10^{-8}~\rm cm/s^2$ \\\hline
        &                                           &                       & \\
      1 & {\sf Systematics generated external to the spacecraft:} &         & \\
        & a) Solar radiation pressure and mass      & $+0.03$               & $\pm 0.01$\\
        & b) Solar wind                             &                       & $ \pm < 10^{-5}$ \\
        & c) Solar corona                           &                       & $ \pm 0.02$ \\
        & d) Electro-magnetic Lorentz forces        &                       & $\pm < 10^{-4}$ \\
        & e) Influence of the Kuiper belt's gravity &                       & $\pm 0.03$\\
        & f) Influence of the Earth orientation     &                       & $\pm 0.001$ \\
        & g) Mechanical and phase stability of DSN antennae       &         & $\pm < 0.001$\\
        & h) Phase stability and clocks             &                       & $\pm <0.001$ \\
        & i) DSN station location                   &                       & $\pm < 10^{-5}$ \\
        & j) Troposphere and ionosphere             &                       & $\pm < 0.001$ \\[10pt]
      2 & {\sf On-board generated systematics:}     &                       & \\
        & a) Radio beam reaction force              & $+1.10$               & $\pm 0.11$ \\
        & b) RTG heat reflected off the craft       & $-0.55$               & $\pm 0.55$ \\
        & c) Differential emissivity of the RTGs    &                       & $\pm 0.85$ \\
        & d) Non-isotropic radiative cooling of the spacecraft    &         & $\pm 0.48$\\
        & e) Expelled Helium produced within the RTGs             & $+0.15$ & $\pm 0.16$ \\
        & f) Gas leakage                            &                       & $\pm 0.56$ \\
        & g) Variation between spacecraft determinations          & $+0.17$ & $\pm 0.17$ \\[10pt]
      3 & {\sf Computational systematics:}          &                       & \\
        & a) Numerical stability of least-squares estimation      &         & $\pm0.02$\\
        & b) Accuracy of consistency/model tests    &                       & $\pm0.13$ \\
        & c) Mismodelling of maneuvers               &                       & $\pm 0.01$ \\
        & d) Mismodelling of the solar corona        &                       & $\pm 0.02$ \\
        & e) Annual/diurnal terms                   &                       & $\pm 0.32$\\[10pt]
 \hline &                                           &                       & \\
        & Estimate of total bias/error              & $+0.90$               & $\pm 1.33$ \\
        &                                           &                       & \\
 \hline\hline
\end{tabular} \end{center} \end{table*}

Table~\ref{tableJPL}, extracted from Ref.\ \cite{JPL}, provides an assessment of the systematic contributions to the acceleration budget; the different orders of magnitude of the various effects involved indicate that these do not account for the reported anomaly; however, in another estimate of the heat dissipation of several spacecraft components, it is claimed that a combination of sources could explain the anomalous acceleration \cite{Lou}. A more recent and complete work has tackled the complex issue of modelling the Pioneer probes with great detail, aiming at the description of all relevant thermal effects with a sufficient accuracy \cite{Toth}; parallel efforts are being pursued by other groups within the Pioneer collaboration team \cite{Lisbon}.

Preliminary results appear to indicate that up to one third of the total magnitude of the reported anomaly may be due to thermal effects \cite{Slava}. However, it should be taken into account that the several modelling strategies and parameter estimations might cloud the overall picture -- so that the physical significance could be difficult to distinguish amongst the technical depth
of the computational reconstitution. For this reason, in Ref. \cite{Lisbon} a complementary methodology is developed, that purposely diverts its attention away from the full modelling of intricate engineering detail, but instead focuses on the physical basis
of the discussed thermal behaviour \cite{Lisbon}. This approach specifically assumes that, although a simplified modelling of specific details might diminish the global confidence of the obtained results, the added simplicity and computational promptness and speed is highly advantageous for a sensitivity analysis of the several relevant parameters.

In this paper, the main features and results obtained so far with this method based on point-like sources are reviewed, including discussion of test cases, inclusion of diffusive reflectivity (and, in the future, specular), a preliminary order of magnitude budget for the thermal contributions from the spacecraft components, and its compatibility with previous studies.

\section{Source Distribution Method}

\subsection{Motivation}

As discussed above, a thorough characterization of the Pioneer anomaly is required before more definitive statements about its origin can be made. This is the rationale behind the task of recovering and analyzing the full flight data, as well as the understanding of the overall thermal behaviour of the Pioneer probes, which can give rise to a non-negligible acceleration. The key physical problem is to ascertain how thermal radiation is emitted and reflected by the external surfaces of the probes, and then calculate the resulting reaction force.

Hence, alongside with the proposed finite elements model, a faster, more versatile approach was put forward \cite{Lisbon}. This methodology, based on a distribution of a few point-like thermal sources, simulates the thermal radiation emitted from the spacecraft and takes into account what fraction is radiated directly into space, or reflected and absorbed by another surface of the spacecraft. 
This approach allows for a faster discrimination of different contributions from the major constituents of the vehicles: as a first approach, the RTGs, antenna dish and main bus compartment.

In favour of this approach, one can state that it is impossible to model the
Pioneer spacecraft in a very precise way: as it was built decades ago, the
accuracy of the blueprints and existing models is limited; furthermore, such a long period in space leaves us with no detailed knowledge of the precise material properties, its degradation or damage. Hence, educated guesses will have to be
taken, setting an intrinsic limit to the accuracy of obtainable results.

Furthermore, the fitting of thermal models to temperature data is also restricted by the data set (stemming from the onboard temperature sensors --- six on the main bus and two on the RTGs) and lack of knowledge regarding the optical properties of the materials also yield uncertainties in the final result. Hence, it is clear that the total electrical power --- which is well known --- must be the fundamental parameter for any analysis. Our approach is based on this principle, besides its computational simplicity and focus on the relevant physics.

Clearly, the distribution of power over the external surfaces of the spacecraft is essential to the quantification of thermal radiation effects. The insulation of the spacecraft walls should limit the related gradient of the temperature along the main external surfaces (except in particular places --- {\it e.g.} the louvers, which can be modeled as separate sources): we argue that small details and gradients in temperature of these external surfaces do not affect the results to a large extent, as shall be demonstrated by an analysis of the overall effect obtained upon variation of the number of point-like sources (keeping the power constant). After considering these assumptions, a sensitivity analysis of the several assumptions may be performed --- including shape modelling, temperature gradients, and total power emitted by each surface; this sensitivity analysis is advantageously achieved for the proposed model, due to its short computation time and straightforward physical basis.

An important simplification relies on the spin stabilization of the Pioneer spacecrafts, as the off-axis reaction forces will cancel out over time. Due to the geometry of the probes, the majority of small contributions possibly not taken into account are expected to be normal to the axis of rotation, and therefore irrelevant. This can be verified through a variation of the radiation distributions, which lead to similar values of the anomalous acceleration, as expected (however, the total reaction torque due to thermal radiation cannot be reliably predicted by the model here discussed).

As hinted above, the simplicity of the formulation allows for a enhanced visibility of the involved physics throughout the entire modelling, thus allowing for a closer scrutiny of the method and derived results. Again, it is emphasized that the primary goal of our approach is to study a wide spectrum of the parameter space relevant for the thermal modelling of the Pioneer probes. This method is clearly not as comprehensive as a finite element model; however, it allows for a direct interpretation of results, higher adaptability and drastically reduced computation times.

Before the key issue of the hypothetical thermal origin of the Pioneer anomaly is addressed, the self-consistency of the method should be strongly established: this is achieved through a battery of test cases that rely on simplistic geometrical arrangements. The validity of the adopted point-like source approach is also verified by a study of the convergence of results with the increase of the number of sources.

\subsection{Physical Formulation}

The method outlined in Ref. \cite{Lisbon} is based on a distribution of isotropic and point-like sources so that, if $W$ is the emitted power, the time-averaged Poynting vector for an isotropic source positioned at $(x_0,y_0,z_0)$ is given by

\beq \mathbf{S}_{\mr{iso}} = {W \over 4\pi}{(x - x_0, y - y_0, z -
z_0) \over \left[(x - x_0)^2 + (y - y_0)^2 + (z - z_0)^2
\right]^{3/2}} ~~. \eeq

\noindent A Lambertian source is characterized by a radiation intensity proportional to the cosine of the angle with the normal unit vector,

\beq \mathbf{S}_{\mr{Lamb}} = {W \cos \th \over \pi}{(x - x_0, y -
y_0, z - z_0) \over \left[(x - x_0)^2 + (y - y_0)^2 + (z - z_0)^2
\right]^{3/2}} ~~. \eeq

\noindent Typically, isotropic sources are used to model point-like
emitters and Lambertian sources to model surfaces. The resulting force and the amount of energy illuminating a surface are obtained by integrating the Poynting vector of the source distribution over this surface. The total energy is given by the time-averaged Poynting vector flux,

\beq E_{\mr{ilum}} = \int \mathbf{S} \cdot \mathbf{n} ~ d A = \int \mathbf{S}(\mathbf{G}(s,t)) \cdot \left({\partial
\mathbf{G} \over \partial s} \times{\partial \mathbf{G} \over
\partial t} \right)~ d s\,d t ~~. \eeq

\noindent where ${\mathbf G(s,t)}$ parameterizes the
surface. Integration of the resulting force -- as given by the radiation pressure multiplied by the unitary normal vector, yields the total force acting upon that
surface. The radiation pressure is given by

\beq \label{prad} p_{\mr{rad}}={\al \over c} \mathbf{S} \cdot \mathbf{n}~~, \eeq

\noindent where a radiation pressure coefficient $1 \leq \alpha \leq 2$ is introduced, so that $\al=1$ indicates full absorption, while $\al < 2$ gives full diffusive reflection.

The force acting on the source of the radiation may be calculated by integrating the radiation pressure multiplied by a normalized radial vector along a generic surface,

\beq \label{emit}\mathbf{F}_{\mr{emit}}=\int {\mathbf{S} \cdot
\mathbf{n} \over c} {\mathbf{S} \over ||\mathbf{S}||} ~ dA ~~. \eeq

In an object endowed with a complex geometry (such as the Pioneer spacecrafts) shadows cast by the surfaces that absorb and reflect the radiation must be taken into account: the magnitude of this shadowing effect is obtained through a similar expression and then subtracted to the force obtained for the emitting surface.
Alternatively, one may use an integration surface that includes the illuminated surfaces: the total result is the sum of all effects, $ \mathbf{a}_{\mr{total}}=\sum_i \mathbf{F}_i / m$.

\subsection{Test cases}

\label{test_cases}

In order to demonstrate the efficiency of the method proposed in Ref. \cite{Lisbon}, a set of test cases is examined. The main question answered through this assessment concerns the ability to adequately represent thermal radiation emitted from an extended surface as an equivalent small number of point-like sources (opposed to considering many small thermal radiating elements). The test cases consider a square emitting surface with $1~\mr{m^2}$, so that the force on the emitting surface, shadow caused by another surface at a given position and radiation pressure on that surface may be computed. The results for different numbers of sources are compared, keeping the total power constant: the result should converge to the
exact solution with the increase of the number of radiation sources. In Ref. \cite{Lisbon} it is shown that a reliable error estimate may be obtained modelling the surface with just a small number of sources.

\begin{figure}

\epsfxsize=8.5cm \epsffile{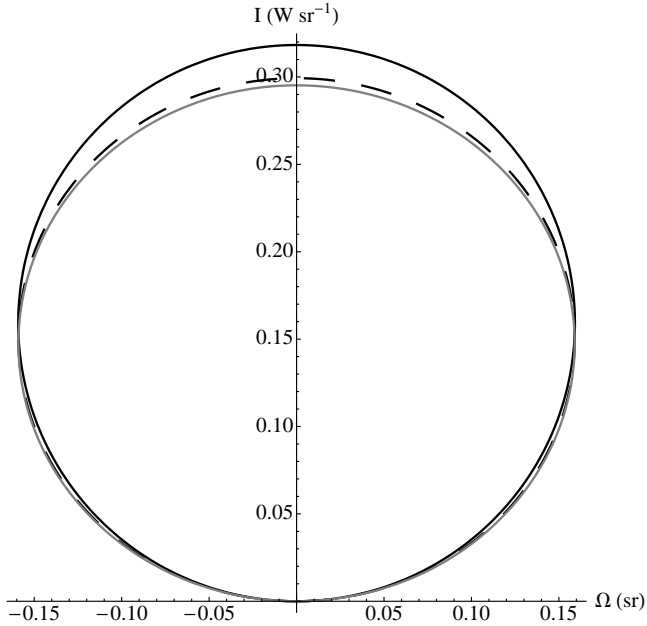} \caption{Polar
plot of the intensity variation with elevation of the radiation
emitted by a surface, when considering 1, 4, 16, 64 or 144
Lambertian sources, maintaining the total emitted power constant.
The intensity diminishes with the number of sources.}

\label{elevation}

\end{figure}

\begin{figure}

\epsfxsize=8.5cm \epsffile{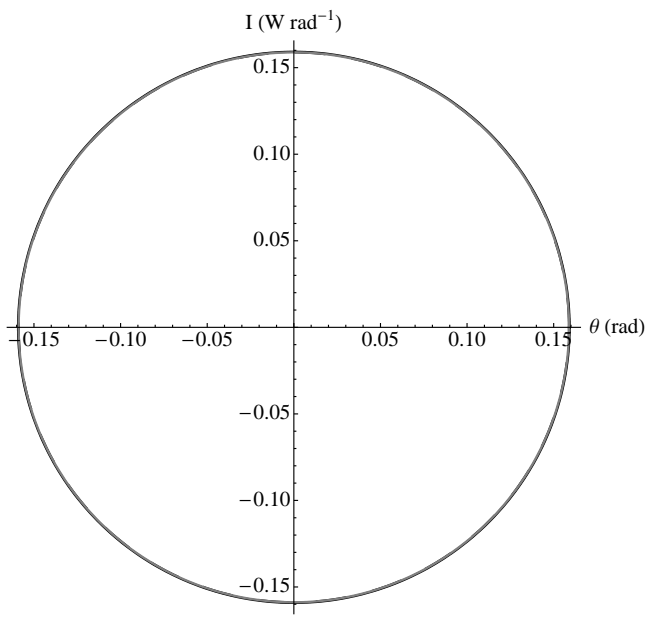} \caption{Same as
Fig.~\ref{elevation} but for intensity variation with azimuth.
Intensity diminishes with the number of sources. Notice that beyond
16 sources the results are almost coincident.}

\label{azimuth}

\end{figure}

Taking a distribution of Lambertian sources on a surface on the $0xy$ plane and using Eq.\ (\ref{emit}) one obtains, for radiation emitting surfaces without any other illuminated surfaces, a force pointing in the $z$-axis and of magnitude $(2/3)
W_{\mr{surf}}/c$, depending only on the total emitted power. However, the calculations of the shadow and pressure radiation on other surfaces do depend on the source distribution: the variation of the radiation intensity with
the elevation and the azimuth for 1, 4, 16, 64 and 144 source
meshes is displayed in Figs. \ref{elevation} and \ref{azimuth}. These plots show that the maximum deviation occurs at the higher elevations and is
less than 10\%, for just one point-like source. Actual deviations for angles relevant to the Pioneer will be considerably smaller. This estimate may be confirmed by calculating the force acting on another $1~\mr{m^2}$ surface, for several different positions: a total of nine configurations were taken, each with different positions and tilt angles (see Table \ref{testcase_configs}); the deviation between the 1, 4, 16, 64 and 144 source meshes is then obtained.


\begin{table}[ht] \begin{center}
\caption{Positions considered for the second surface in test cases; distances between both surfaces are typical for the Pioneer spacecraft.}

\label{testcase_configs}


\begin{tabular}{cccc}

   \hline\hline
   Test case ~&~ Surface Centre Position ~&~ Surface Tilt Angle  \\
   \# & ($\mr{m}$) & ($^\circ$) \\
   \hline
   $1$ & ($0$, $2$, $0.5$) & $90$  \\
   $2$ & ($2$, $0$, $1.5$) & $0$   \\
   $3$ & ($2$, $0$, $1.5$) & $30$  \\
   $4$ & ($2$, $0$, $1.5$) & $60$  \\
   $5$ & ($2$, $0$, $1.5$) & $90$  \\
   $6$ & ($1$, $0$, $2$)   & $0$   \\
   $7$ & ($1$, $0$, $2$)   & $30$  \\
   $8$ & ($1$, $0$, $2$)   & $60$  \\
   $9$ & ($1$, $0$, $2$)   & $90$  \\
   \hline\hline

\end{tabular} \end{center} \end{table}

\begin{figure}

\epsfxsize=4.5cm \epsffile{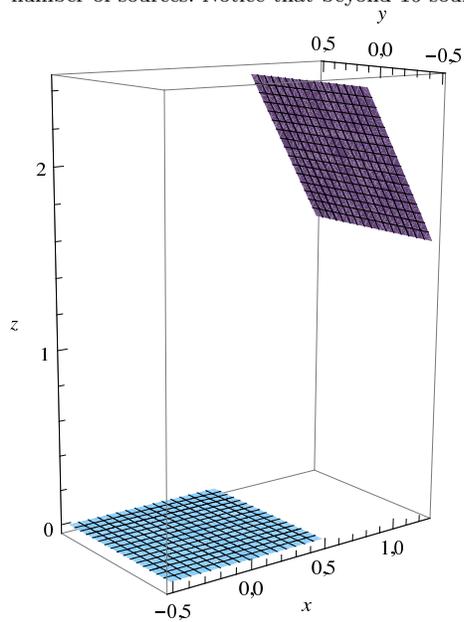} \caption{Geometry of
test case~8 (cf.\ Table~\ref{testcase_configs}): thermal emission
from a surface is simulated by a varying number of Lambertian
sources evenly distributed on the surface, keeping the total
emitted power fixed, and the effect on another surface is
observed. This test case leads to the highest variation of results with the
number os sources.}

\label{testfig8}

\end{figure}

The highest deviation was obtained for test case~8, confirming the assumed scenario --- since the second surface is at a high elevation
from the emitting surface, as shown in Fig.\ \ref{testfig8}. The
results in Table~\ref{table8} indicate a difference of about
6\% between the force obtained with one source and finer meshes (16, 64 and 144 sources). The latter are within 0.5\% of each other, and the intermediate
4 source mesh has a deviation of just 1.5\%.


\begin{table}[ht] \begin{center} \caption{Results for test case~8 (cf.\ Table~\ref{testcase_configs})
considering a total emission power of $1~\mr{kW}$. The force components due to shadowing on the secondary surface are almost independent of the number of
sources on the emitting surface.}

\label{table8}


\begin{tabular}{cccc}

   \hline\hline
   Sources ~&~ Energy Flux ~&~ Force components $(x,y,z)$ \\
   \# & ($\mr{W}$) & ($10^{-7}~\mr{N}$) \\
   \hline
   $1$   & $45.53$ & ($2.016 $, 0, $2.083 $)  \\
   $4$   & $45.53$ & ($1.918 $, 0, $2.003 $)  \\
   $16$  & $45.53$ & ($1.895 $, 0, $1.984 $)  \\
   $64$  & $45.53$ & ($1.890 $, 0, $1.979 $)  \\
   $144$ & $45.53$ & ($1.889 $, 0, $1.978 $)  \\
   \hline\hline

\end{tabular} \end{center} \end{table}

\begin{figure}

\epsfxsize=8.5cm \epsffile{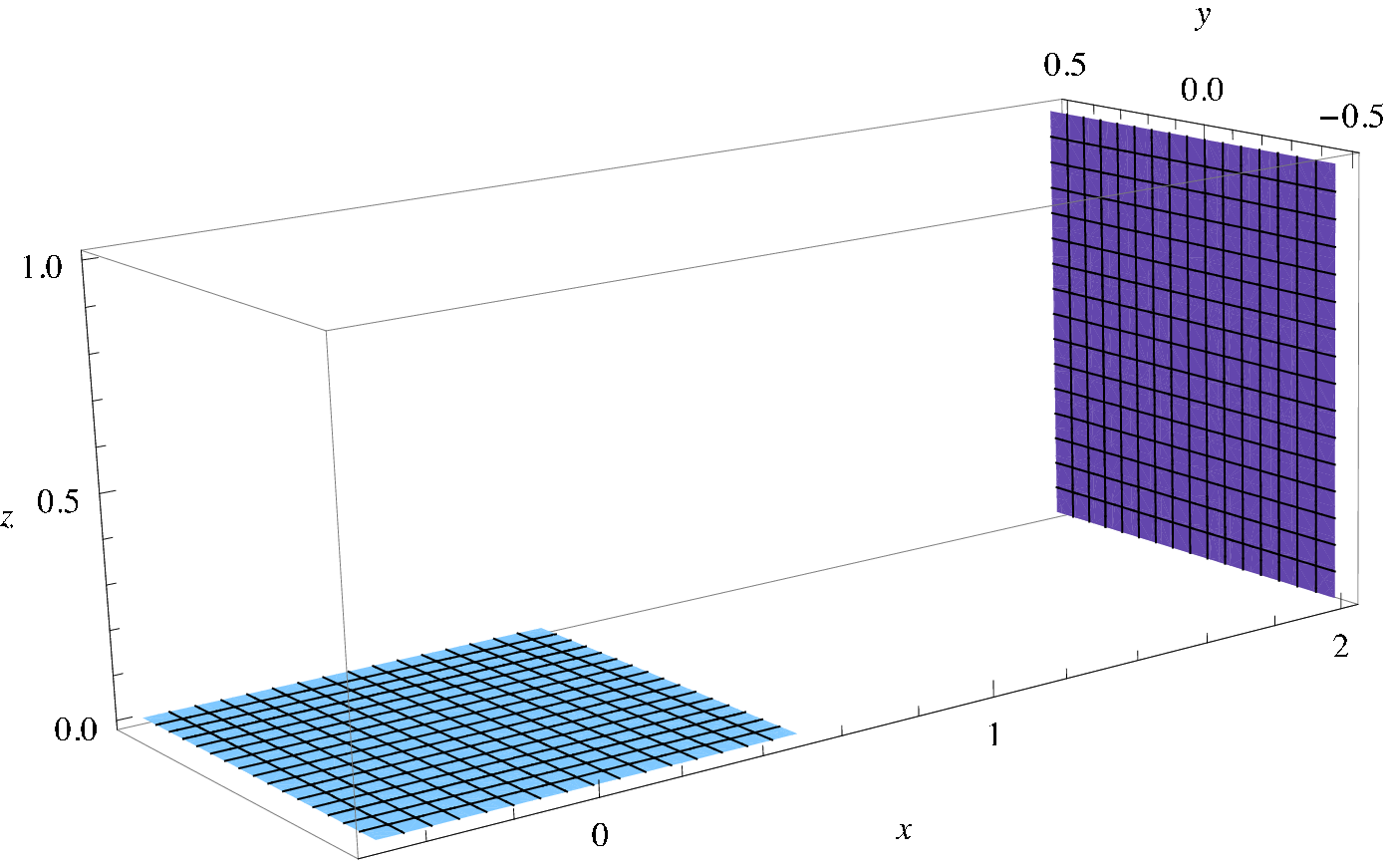} \caption{Same as
Fig.~\ref{testfig8} for test case~1.}

\label{testfig1}

\end{figure}

Test cases~1 and 3 are the most relevant, since these depict configurations typical of the Pioneer spacecraft: the first case yields results present in Table~\ref{table1} (see also Fig.~\ref{testfig1}), showing that, for 16, 64 and 144 sources, variation in the energy flux and force is smaller than 0.5\%. Furthermore, comparison with finer meshes shows that results vary by less than 5\% for one source, and less than 1.5\% for a four source mesh. For test case 3, the results in Table \ref{table2} show a variation of less than 5\% between the results for one and 144 sources. As in the previous cases, convergence is achieved for the 16, 64 and 144
source meshes, with a variation of less than 0.25\%.


\begin{table}[ht] \begin{center} \caption{Same as Table~\ref{table8} for test case~1.}

\label{table1}


\begin{tabular}{cccc}

   \hline\hline
   Sources ~&~ Energy Flux ~&~ Force components $(x,y,z)$ \\
   \# & ($\mr{W}$) & ($10^{-7}~\mr{N}$) \\
   \hline
   $1$   & $15.34$ & ($0.9300 $, 0, $0.1514 $)  \\
   $4$   & $15.92$ & ($1.028 $, 0, $0.1638 $)  \\
   $16$  & $16.09$ & ($1.038 $, 0, $0.1675 $)  \\
   $64$  & $16.13$ & ($1.040 $, 0, $0.1684 $)  \\
   $144$ & $16.14$ & ($1.041 $, 0, $0.1686 $)  \\
   \hline\hline

\end{tabular} \end{center} \end{table}



\begin{table}[ht] \begin{center} \caption{Same as Table~\ref{table8} for test case~2.}

\label{table2}


\begin{tabular}{cccc}

   \hline\hline
   Sources ~&~ Energy Flux ~&~ Force components $(x,y,z)$ \\
   \#    & ($\mr{W}$)   & ($10^{-7}~\mr{N}$) \\
   \hline
   $1$   & $19.20$ & ($0.4952 $, 0, $1.037 $)  \\
   $4$   & $19.83$ & ($0.5032 $, 0, $1.082 $)  \\
   $16$  & $19.99$ & ($0.5050 $, 0, $1.093 $)  \\
   $64$  & $20.03$ & ($0.5054 $, 0, $1.096 $)  \\
   $144$ & $20.04$ & ($0.5055 $, 0, $1.096 $)  \\
   \hline\hline

\end{tabular} \end{center} \end{table}


Ultimately, one concludes that a mesh with four Lambertian sources, which shows deviations of about 1.5\%, provides the desired balance between precision and simplicity; this yields a convenient illustration of the ability of the developed method to estimate the radiation effects on the Pioneer probes. Given that the deviations are always well below 10\%, even when the roughest approximations are considered, it appears that the application of this source distribution method to the scales and geometry involved in the Pioneer anomaly problem yields consistent and convergent estimates of the thermal radiation effects.

\section{Thermal Radiation Model of the Pioneer Spacecraft}

\subsection{Geometry}

\begin{figure}

\epsfxsize=6cm \epsffile{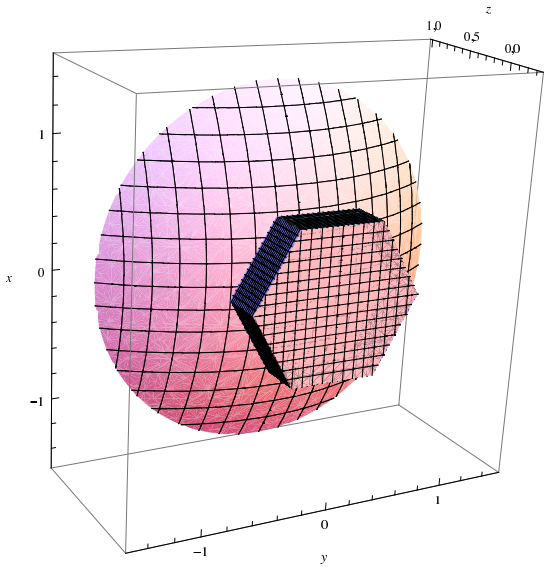} \caption{Pioneer
spacecraft model geometry, back view:
high gain parabolic antenna and hexagonal main bus compartment.}

\label{pioneer}

\end{figure}

A simplified model of the Pioneer spacecraft can be obtained by  considering some plausible, physically motivated hypotheses, such as the aforementioned probes' spin stabilization, and also the assumption of steady-state thermal equilibrium throughout most of their journey. Also, a simplified version of the spacecraft
geometry may be considered as a first approach to the problem of estimating its thermal radiation: hence, one considers only its most important features (as illustrated in Fig.~\ref{pioneer}): the parabolic antenna, RTGs and a prismatic equipment compartment, with dimensions derived from the available Pioneer technical drawings.
The estimation of thermal effects proceeds by integrating the emissions of the RTG and lateral walls of the equipment compartment along the visible portion of the antenna; since the emissions stemming from the front surface of the Pioneer probes is never reflected, this element is considered as a whole. Furthermore, one may discard the surface of the compartment that faces the antenna, since its contribution to the anomalous acceleration is negligible for obvious geometric reasons: its emitted radiation, which is mainly radial, is attenuated through multiple reflections. Finally, the antenna should have a very low, approximately uniform temperature (as discussed in Ref. \cite{Slava}), also yielding a negligible effect.

It will be demonstrated how this simplified model is able to reproduce the most important
contributions to the thermal reaction force: the RTGs and the main
equipment compartment are effectively responsible for the overwhelming part of the
emitted thermal radiation. In the latter case, most of the effect arises from the louvers located in the front wall (facing away from the Sun) -- with consequences for the total power distribution.

\subsection{Point-like Source Distribution}

\label{source_distrib}

An analysis of each of the three main contributions  to the total thermal effects is now developed. The front wall of
the spacecraft, where an axial force with magnitude $(2/3)
W_{\mr{front}}/c$. The side walls of the main compartment is calculated by integrating the shadow and radiation pressure components along the antenna. One may neglect the shadow of the RTGs, due to their smallness and relatively distant positioning, which would yield a contribution mostly in the radial direction.

Following the approach depicted in the test cases already discussed, the integration is performed with an increasing number of sources, thus allowing for evaluation of the method's convergence: this occurs rather quickly, with deviations consistently below 2.5\%. The values obtained indicate that 16.8\% to 17.3\% of the power emitted from the side walls of this compartment is converted into approximately Sun-ward thrust along the $z$-axis.

The relevance of assuming a non-uniform temperature distribution is also assessed, by varying the relative power of the point-like sources in each surface,
keeping the total surface power constant. A variation of 20\% in power between sources, which models a 5\% temperature variation, does not produce significant changes in the final result --- since observed relative differences are smaller than 1\%.

Finally, the RTG contribution is computed through two different
models. The first, simpler scenario, mimics each RTG with a single
isotropic source. In this case, the point-like source has the whole
power of the RTG. In the second model, the cylindric shape of the
RTG is taken into account and a Lambertian source is placed at each
base. Actually, it is only necessary to consider the source facing
towards the centre of the spacecraft, as the remaining RTG radiation
will be emitted radially and its time-averaged contribution
vanishes. In this case, the Lambertian source has a certain amount
of the total RTG power, as discussed in the following Sections.
Depending on the model considered, either 1.9\% of the total power
or 12.7\% of the power emitted from the base of the
cylinder (equivalent to approx.\ 2\% of total RTG power, if temperature is uniform) is
converted into thrust. 

In the following section, more accurate figures will be obtained, which include the diffusive reflection included in Eq. (\ref{prad}).

\subsection{Available Power}

Amongst the panoply of physically meaningful parameters, the available power on the Pioneer spacecraft is one of the few that is reasonably well known. For this reason, the methodology pursued here takes this quantity as  the independent variable from which all
estimates of the resulting thermal effects are derived, instead of
temperature readings -- which must be interpolated from the isolated sensor sites to the remainder elements of the spacecraft, an objective outside of the scope of this study. Clearly, the (assumed uniform) temperature of a surface $i$ is related to the energy balance to the spacecraft in steady-state conditions, through Stefan's law,

\beq \dot{E}_{\mr{absorb}}+\dot{E}_{\mr{gen}}=\sum_i A_i \epsilon_i
\sigma T_i^4~~, \eeq

\noindent where $A_i$ are the relevant areas and $\ep_i$ the
emissivity of surface $i$.

Since one does not have accurate information about the optical properties of the surfaces, as well as their evolution in time, the temperature estimates
are quite uncertain. Also, note that a variation in the emissivity requires the computation of a new solution for the temperature distribution, in order to obtain the correct total power and maintain energy conservation.

The power generated onboard the probes comes from the two
Plutonium-238 powered RTGs, with just a fraction of the generated heat
converted into electricity: the remaining power is dissipated as
heat, which will be transfered to the central compartment through conduction in the truss assembly. However, given the small section
of this structure, one may disregard this contribution and take the RTGs as isolated elements, with all produced thermal power radiated directly from them.

A considerable proportion of the electrical power is consumed by the high gain antenna, with the remainder being used by the various instruments located
in the main compartment. As found in Ref.\ \cite{JPL}, the total RTG thermal power at launch amounts to 2580~W, yielding 160~W of electrical power -- so that, at
launch, approximately 2420~W of thermal power was radiated away by
the RTGs. Considering the plutonium decay with a half-life of 87.74 years, the total on-board power variation with time (in years) is given by

\beq W_{\mr{tot}}(t)= 2580 \exp \left(-{\frac{t \ln 2
}{87.74}}\right)~~. \eeq

\noindent However, telemetry data reveals that the electrical power decayed at a faster rate: the electrical heat in the body of the spacecraft was around 120~W at launch, dropping to less than 60~W at the latest stages of the mission \cite{Toth}, following an approximate exponential decay with a half-life of only 24 years. This accelerated power decay reflects the added effect of thermocouple degradation, which reduces the conversion efficiency of the RTGs.

\section{Results and Discussion}

\subsection{Order of Magnitude Analysis}

Before proceeding to a more rigorous numerical estimate, the results detailed in the previous Sections can be used to perform a preliminary order of
magnitude analysis; this will serve to produce a concrete figure of merit for the overall acceleration arising from thermal effects, which can then be compared with the $a_\mr{P} \sim 10^{-9}~\mr{m/s^2}$ scale of the Pioneer anomaly.

The spacecraft specifications indicate a total mass for the probes $m_{Pio} \sim
230~\mr{kg}$, and RTG and equipment compartment powers
$W_{\mr{RTG}} \sim 2 ~\mr{kW}$ and $W_{\mr{equip}} \sim 100
~\mr{W}$, respectively. As discussed, integrating the corresponding thermal emissions yields the proportion of emitted power that is effectively converted into thrust. Assuming a uniform temperature and emissivity in the RTGs and main equipment compartment, which leads to a power emitted from each surface proportional to its area; with the simple model discussed is Section \ref{source_distrib}, this yields

\beqa F_{\mr{RTG}} \sim 2 \times 10^{-2} {W_{\mr{RTG}} \over c},\\
\nonumber F_{\mr{sides}} \sim 10^{-1} {W_{\mr{equip}} \over
c}~~,~~F_{\mr{front}} \sim 2 \times 10^{-1} {W_{\mr{equip}} \over
c}~~. \eeqa

The acceleration of the spacecraft due to the thermal effects arising from the power dissipation of the RTGs and equipment compartment is easily obtained,

\beqa a_{\mr{RTG}} & \sim & 2 \times 10^{-2} {W_{\mr{RTG}} \over m_{Pio}c}
\sim 6 \times 10^{-10}~\mr{m/s^2}~~,\\ \nonumber a_{\mr{equip}} &
\sim & 1.5 \times 10^{-1} {W_{\mr{equip}} \over m_{Pio}c} \sim 2.2 \times
10^{-10}~\mr{m/s^2}~~, \eeqa

\noindent so that both contributions are relevant when estimating the acceleration induced by thermal radiation from the Pioneer probes: since the RTGs and the instrument compartment produce similar thermal effects, one cannot rely solely on one of these sources, confirming the previous discussion in Refs. \cite{Lou,Toth}).

\subsection{Thermal Force Estimate}

Having gained momentum through the estimate depicted above, a more thorough evaluation of the existing thermal effects using the point-like source modelling developed in Ref. \cite{Lisbon} follows: to this effect, one considers a model with 4 sources per side panel of the equipment compartment, and Lambertian sources at the bases of the RTGs (see discussion in section \ref{source_distrib}) -- the configuration that provides what is deemed as the best compromise between accuracy and computation time, with the aforementioned deviation $< 0.5 \%$ from results using finer meshes. The acceleration is obtained from the axial component of the integration of the radiative emissions of the Pioneer probes,

\beq  \label{acceleration} a_{\mr{Pio}} = { \left( 0.168 W_{\mr{sides}} + {2 \over 3} W_{\mr{front}} +
0.128 W_{\mr{base}} \right) \over m_{\mr{Pio}}c}~~, \eeq

\noindent where $W_{\mr{sides}}$ and $W_{\mr{front}}$ are the powers emitted
from the side panels and front of the equipment compartment and
$W_{\mr{base}}$ is the power emitted from the base of the RTG facing the centre of the spacecraft.

To ascribe values to each of these powers, one takes readings from 1998, as found in Ref. \cite{Toth}, namely the dissipated thermal powers at the RTG and equipment compartment $W_{RTG}=2050 ~\mr{W}$ and $W_{\mr{equip}}=58 ~\mr{W}$, respectively. With the simplest scenario of uniform temperature and
optical properties, this gives

\beqa && W_{\mr{sides}}= 21.75 ~ \mr{W}~~,~~W_{\mr{front}}=18.12~ \mr{W} \\ \nonumber && W_{\mr{base}}=41.11~ \mr{W}~~,\eeqa 

\noindent leading to an acceleration $a_{\mr{Th}}=3.05 \times 10^{-10}~\mr{m/s}^2$. This 
 amounts to about 35 \% of the anomalous acceleration.

From the available temperature maps used in Refs. \cite{Toth,Slava}, one can see that the temperature anisotropies along the sides of the equipment compartment rest within the limits assumed in the test cases, as discussed in Section \ref{source_distrib}. However, the RTGs display significant temperature variation between the wall of the cylinder, the bases and the fins. Also, the front wall of the equipment compartment is expected to produce a larger contribution than that of the side walls, due to the presence of the louvers.

\begin{figure}

\epsfxsize=7cm \epsffile{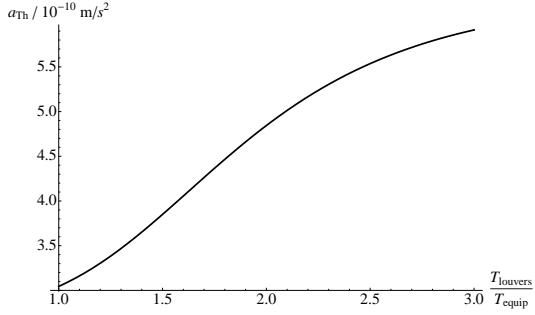} \caption{Variation of the resulting acceleration with the temperature ratio between the louvers and the equipment platform, considering similar emissivities for both multi-layer insulations.}

\label{graphTlouvers}

\end{figure}

\begin{figure}

\epsfxsize=7cm \epsffile{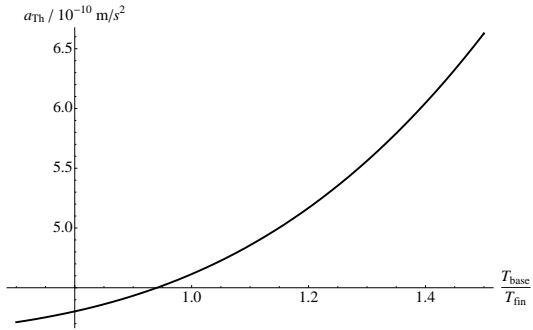} \caption{Variation of the resulting acceleration with the temperature ratio between the base of the RTG cylinder and the fin temperature.}

\label{graphTrtg}

\end{figure}

This said, one can ascertain how does the emitted power in the louvers and at the base of the RTG vary --- two critical parameters in this calculation. If the louvers are closed and assumed to have a emissivity similar to the equipment platform, the variation of the acceleration with the temperature ratio between the louvers and the mean temperature of the platform may be plotted (again, with the total power fixed), as shown in Fig. \ref{graphTlouvers}. The same analysis can be performed with respect to the RTGs, now considering the ratio between the temperatures at the base of the cylinder and the fins (shown in Fig. \ref{graphTrtg}).


Figs. \ref{graphTlouvers} and \ref{graphTrtg} help to better grasp the main advantage of the developed method \cite{Lisbon}: the possibility of conducting a sensitivity analysis that is both fairly quick and sufficiently accurate; from Eq. (\ref{acceleration}) and a plausible variation of the power parameters, temperature readings may be adjusted and different optical properties suitably discussed and inferred.
 
Another scenario that may be explored assumes that the RTG cylinder bases and wall possess a temperature 15\% and 30\% higher than that of the fins, respectively, while the closed louvers have double the temperature of the equipment compartment (with constant emissivity). This yields the following estimates for the power,

\beqa W_{\mr{sides}}= 9.97~\mr{W}~~,~~ W_{\mr{front}}=39.71~\mr{W} ~~,\\ \nonumber
W_{\mr{base}}=49.67~\mr{W}~~, \eeqa

\noindent thus accounting for 57\% of the anomalous acceleration, that is, $a_{\mr{Th}}=5.00 \times 10^{-10}~\mr{m/s}^2$.

Up to this point, the study has considered only the case of full absorption of the radiation by the illuminated surfaces; to counteract this, one includes diffusive reflection by assigning a non-unitary value to the $\al$ parameter in Eq.\ (\ref{prad}). For the relevant wavelengths, the reflectivity of the type of aluminum used in the high gain antenna is commonly around 80\% , yielding  $\al=1.8$; also, the multi-layer insulation of the equipment platform may be modeled by a value $\al=1.7$. Given these conditions and the re-derived illumination factors in Eq.\ (\ref{acceleration}), the same temperature conditions considered in the previous case result in a thermally induced acceleration $a_{\mr{Th}}=5.75 \times 10^{-10}~\mr{m/s}^2$.

Besides allowing for the computation of observables, the derivations put forward in this Section assess the variations occurring when considering different parameters and hypotheses. The three discussed scenarios serve to show that the method developed in Ref. \cite{Lisbon} is well suited to identify sensitive parameters and promptly evaluate the propagation of the existing uncertainties.

\section{Conclusions}

In this contribution we discuss the applicability of the method developed in Ref. \cite{Lisbon} to account for the acceleration of the Pioneer spacecrafts arising from thermal effects, based on point-like sources. After identifying the main power contributions of the various components of the spacecraft, 35\% to 57\% of the reported anomalous acceleration is obtained.

The discussed method displays a reasonable degree of accuracy, with a direct numerical computation error of the order of $10^{-14}$ or less, and deviations due to the approximation of the geometry with point-like sources falling below 1\% (as argued in Sections \ref{test_cases} and \ref{source_distrib}). However, it should be emphasized that this is not a direct indication of the accuracy of the actual thermally induced acceleration, when compared to the reported case of the Pioneer anomaly --- but a measure of self-consistency of the developed method. The aim here is to demonstrate the reliability of the procedure, which should be extended in order to better simulate the physical system of the Pioneer spacecrafts -- while keeping the flexibility and computational speed that have motivated this complementary approach.

Thus, future refinements will focus on an enhanced geometrical modelling, including specular reflection. Furthermore, it will be of great interest to identify the parameters that most directly affect the final result, {\it e.g.} temperatures, emissivities and reflectivities of the various components. Clearly, the encouraging results of the analysis performed in Ref.\ \cite{Lisbon} are intimately connected to the level of agreement with thermal models based on finite element methods: as this work demonstrated, our estimates are so far consistent with these approaches.

\begin{acknowledgements}
This work is partially supported by the Programa Dinamizador de Ci\^{e}ncia e Tecnologia do Espa\c{c}o of the Funda\c{c}\~{a}o para a Ci\^{e}ncia e Tecnologia (FCT, Portuguese Agency), under the project PDCTE/FNU/50415/2003. The work of JP is sponsored by the FCT under the grant BPD 23287/2005.
\end{acknowledgements}


\end{document}